# Beyond Traditional DTN Routing: Social Networks for Opportunistic Communication

Mary R. Schurgot*[1], Cristina Comaniciu*, and Katia Jaffrès-Runser*†

*Stevens Institute of Technology
Hoboken, NJ, USA
†University of Toulouse,
IRIT-INPT ENSEEIHT, Toulouse, France

*Abstract*— This article examines the evolution of routing protocols for intermittently connected ad hoc networks and discusses the trend toward social-based routing protocols. A survey of current routing solutions is presented, where routing protocols for opportunistic networks are classified based on the network graph employed. The need to capture performance tradeoffs from a multi-objective perspective is highlighted.

## Introduction

Delay-tolerant networks (DTNs) [1] are partitioned wireless ad hoc networks with intermittent connectivity. Additional terminology in this family of dynamic networks includes *disruption-tolerant networks*, *intermittently connected networks*, and *opportunistic networks*. DTNs are never fully connected at any point in time, but points of disconnection **may** be predictable as in vehicular networks following transportation schedules or networks with satellites traversing orbits [2]. In an intermittently connected network (ICMAN) or an opportunistic network, nodes rarely have information on the changing network topology [3][4]. Nodes may not know the availability of future encounters, but the network may benefit from learning such patterns over time. Thus, subsets of nodes in transmission range leverage cooperation during pairwise contacts to forward data towards a destination [4].

The designers of these dynamic networks often rely on the mobility of nodes to route messages and bridge partitions. Intermediate relays may be required to store messages and deliver them to destinations as they are encountered, i.e. enter into radio range. Mobility-assisted routing in DTNs is enabled by this "store-carry-forward" paradigm. A variant of this approach is the store-carry-replicate strategy, which replicates the routed packets, thus increasing the number of copies in the network. As investigated in [2], traditional ad hoc routing protocols must be adapted within a DTN architecture. Classical proactive or reactive routing approaches proposed for regular ad hoc networks do not work for these challenged DTNs, due to the fact that an end-to-end path may not be available at the time of transmission. However, over time, as different links come up and down thanks to mobility (or other environmental characteristics), the dynamic evolution of connectivity graphs over a longer time interval may lead to an asynchronous end-to-end path.

Existing DTN routing protocols evolved from enabling the transfer of any amount of data to carefully selecting intermediate nodes to efficiently carry information. Forwarding schemes were adapted over time to address different performance measures: delivery ratio, message latency, and overhead. The design of DTN routing algorithms may be application-specific, but generally all schemes should balance the overhead from redundant copies with successful delivery and minimal delay. In this work we emphasize the need for multi-objective optimization to better understand performance tradeoffs in opportunistic networks.

Improved performance amounts to identifying suitable carriers for a specific destination. Nodes may be drawn to particular geographic regions or influenced by the behavior of other nodes. With an underlying assumption that the mobility process is ergodic and stationary, algorithms have been designed to predict the future from past behavior. This assumption may not always be valid and slower changing attributes, like social connections, may be leveraged to enable efficient message delivery. Social relationships are expected to vary slower than the transmission links between mobile nodes [5]. In fact, the application of social network theory to model delay-tolerant networks has led to the design of a new class of routing solutions. Forwarding algorithms like SimBetTS [6] and BUBBLE [5] consider a node's role in the social structure of the network to make routing decisions.

Social-based protocols may quantify the social network structure, identify socially-similar nodes, and/or utilize context information [4] like shared interests or community affiliations. Social-based routing is a particularly relevant solution for opportunistic networks with a social component like pocket-switched [5] and mobile peer-to-peer networks [7].

In this work, we present the evolution of DTN routing protocols and highlight the application of social network theory to communication systems. Previous tutorials and surveys focused on formally defining a DTN architecture and discussing routing solutions. Unique to our review is the classification of routing protocols based on the network graphs we define: the dynamic wireless graph composed of every available link in time; the contact graph calculated from the aggregation of past wireless links; and finally the social graph formed by interpersonal relationships. The intent of this article is not to provide a comprehensive review of all DTN protocols. Instead, we chose a cross section of protocols that chronicles the development of sophisticated routing solutions for intermittently connected ad hoc networks. We begin with a description of the wireless graph and associated protocols; then transition to the contact graph. The social graph is then introduced and defined from two perspectives. The article concludes with a discussion of open research issues and challenges for the application of social networking for opportunistic communication.

---

[1] The first author is now with LGS Bell Labs Innovations, Florham Park, NJ, USA. Email: maschurg@lgsinnovations.com

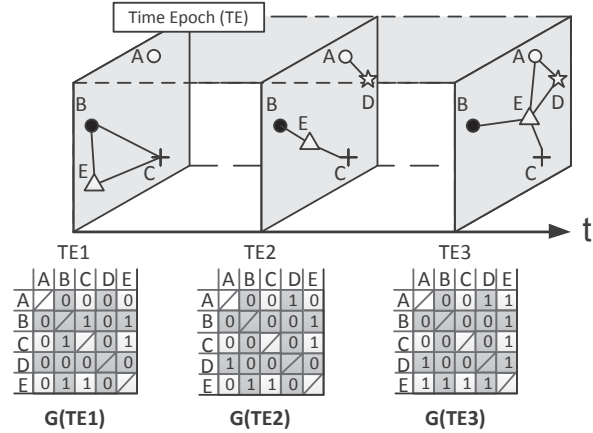

Fig. 1. Node connections over time in a DTN. Matrix G(TE) represents the wireless graph G of the network at time epoch TE.

## THE WIRELESS GRAPH FOR DATA TRANSFER

Routing solutions rely on the existence of wireless links between nodes. In the networks of interest, these links are not persistent in time. The network is typically sparse and the topology can change frequently. Thus, we need a three-dimensional graph, the wireless graph, to represent the network at each time epoch. A new time epoch $TE$ begins when a change to the topology takes place. The wireless graph is an instantiation of a time-varying graph, and a change in state is captured by a new time epoch. The wireless graph is a dynamic undirected graph with an edge between nodes signifying the presence of a wireless link in both directions. Information may not be known on the exact quality of the links, just that the nodes are within range of the radio transceivers and the channel can support communication at a minimum rate. A value of 1 in the connectivity matrix $G(TE)$ indicates the presence of a link and 0 otherwise. Each time a neighboring node moves in or out of transmission range, the wireless graph and associated matrix change.

An illustrative example of a wireless graph in Fig. 1 shows the state of a DTN for three time epochs. New links become available over time and form an asynchronous end-to-end path between nodes $A$ and $C$. The network is fully connected in time epoch 3 ($TE3$) due to the nodes' mobility and due to the availability of a highly central node $E$. As a consequence, node $A$ must wait until $TE3$ to send its message or any message it has to relay to $C$. Legacy ad hoc protocols were not designed to support this type of communication. Delay-tolerant routing, however, makes a forwarding decision at each encounter instead of identifying a fixed route at the onset.

The protocol employed by a network determines path selection and thus sets network performance. For the direct delivery case, $A$ and $C$ may eventually (or potentially never) be in transmission range. Direct delivery from a source to a destination sets the upper bound on delay. Without deterministic knowledge of future node encounters, the fastest path is identified through flooding all nodes in contact at each time epoch of the wireless graph. However, flooding necessitates infinite buffer capacity which is of course not tractable in practice. Using a flooding algorithm leads in practice to overloaded buffers for frequently used relays, which in turn leads to dropped packets and consequently poor delivery ratio performance.

As illustrated by the example of Fig. 1, the focus of routing protocols in disconnected networks is to utilize pairwise contacts to enable opportunistic communication. The questions that arise with this approach are: *to whom to forward* and *how much to replicate?* A significant amount of literature exists trying to basically address these questions by proposing various routing solutions. Of these, three benchmark protocols stand out and are used for performance comparisons by almost all more recently proposed protocols: Epidemic [8], Spray and Wait [9], and discussed in the next section, PRoPHET [10]. In the remainder of this section, we present Epidemic and Spray and Wait, which only use information from the wireless graph for routing.

### Epidemic

The Epidemic protocol [8] is based on general broadcasting of messages: nodes freely replicate messages on each encounter until a message has reached a predefined maximum hop count. Messages are not exchanged if a copy is already present in the peer's buffer. Because it is essentially a flooding protocol, Epidemic was shown to have a good packet delivery ratio, but it suffers from very high overhead given the large number of packet copies flooding the network. Although buffer congestion issues have not been addressed in the protocol's design, the authors empirically investigate the impact of buffer size on successful delivery.

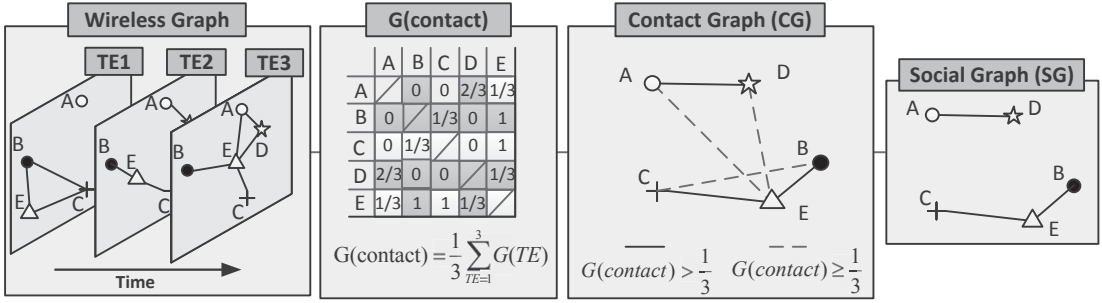

Fig. 2. Network graph classifications.

*Spray and Wait*

The Spray and Wait protocol outperforms all schemes discussed in [9] including Epidemic for a large range of network connectivity scenarios. It is shown to perform close to the optimal oracle scheme (which has complete knowledge of future node encounters, i.e. future states of the wireless graph) for a random waypoint mobility model. The algorithm consists of two phases: spray and wait. During the spray phase, L packet copies are "sprayed" to relays in the network. Then these carriers enter the wait phase until they meet the destination and the message is delivered. Spray and Wait is further defined by the type of spraying employed. With source spray, the source replicates a message to the first L nodes contacted. In binary spray the source keeps $\lceil L/2 \rceil$ copies and distributes the remaining copies to the first node encountered. The relay carries $\lfloor L/2 \rfloor$ copies. This distribution continues recursively for each encounter until each node is left with one copy (the number of copies in the network is kept to L). The optimal number of copies L* is also derived for a specific delay requirement.

THE CONTACT GRAPH FOR EFFICIENT FORWARDING

Clearly, all protocols operate over the underlying wireless graph, and it is how this information is processed which differentiates solutions. As the amount of knowledge available to the protocol increases, network performance with respect to average delay and delivery ratio improves [2]. The algorithms examined by Jain et al. range from extremely simple as with first contact, which forwards a message to the first node encountered, to the fully formulated linear program with information on the wireless graph at each time epoch, the occupancy at each queue, and the traffic demand from each node [2]. Of course complete global knowledge of the wireless graph is not realistic in practice, and the protocols presented here do not possess deterministic information on future connectivity. The contact graph aims to predict these future encounters.

The contact graph is calculated based on aggregating statistics from the dynamic wireless graph. The contact graph serves two purposes:

1) To predict future encounters from statistics of the wireless graph by assuming the mobility process is ergodic and stationary.
2) To reduce the amount of information stored and processed by nodes. By aggregating the data, a node does not need to store a snapshot of the network at each past time epoch.

Entries in the connectivity matrix $G(contact)$ are no longer binary as in the wireless graph. Edge weights are between $0$ and $1$. These weights are calculated during an aggregation window composed of a series of time epochs. A new contact graph ($CG$) and $G(contact)$ can be built for each time window. The contact graph definition and weight assignments depend on the routing solution. For example in [5], edge weights are assigned based on the number of contacts and duration of contacts. The weights of $CG$ in the example of Fig. 2 are set by averaging the node meetings logged in the connectivity matrices $G(TE)$ over three time epochs.

The efficiency of routing protocols which use the contact graph $CG$ is completely dependent on the edge weights and the implemented forwarding rule. A possible forwarding rule using $CG$ in Fig. 2 may be to forward a copy if $G(contact) > \frac{1}{3}$. This is restrictive since one third of the time $E$ can relay messages from $A$ to $C$. The aggregation threshold could be adapted to choose $G(contact) \geq \frac{1}{3}$; however, this will consume additional resources. Although very simple, this example illustrates well the sensitivity of the network performance to the fine tuning of the model and routing decisions, as well as the tradeoffs involved among various performance metrics.

In the remainder of this section, we discuss PRoPHET, MaxProp, and RAPID, which use information from the contact graph to make routing decisions.

*PRoPHET*

The PRoPHET algorithm [10] studies pairwise contacts to make routing decisions. PRoPHET reduces the overhead by calculating a node's delivery predictability for a specific destination. If an encountered node $B$ has a higher delivery predictability for a given message, carrier $A$ transmits a copy to $B$. The delivery predictability for a node $A$ is based on the number of encounters of $A$, the age of these encounters, and the existence of a transitive property for mutually encountered nodes. PRoPHET was shown to perform better than Epidemic for the community-based scenario and comparable to Epidemic for the random mobility case. Although PRoPHET does not explicitly define a contact graph, the delivery predictability is a metric calculated from the aggregation of the wireless graph over time and thus it fits well within our contact graph based framework.

*MaxProp and RAPID*

The efficiency of routing protocols in DTNs continues to improve upon the performance of these benchmark protocols. Protocols such as MaxProp [11] and RAPID [12] have been demonstrated on vehicular networks with intermittent connectivity and add more realistic constraints on fixed storage space.

In an effort to increase the delivery rate and reduce latency, the MaxProp protocol prioritizes buffered packets for retransmission. Packets with lower hop counts are given priority in order to facilitate quick propagation through the network. Once packets exceed the hop count threshold, packet prioritization is determined by the probability that two peers meet calculated using incremental averaging [11]. Acknowledgements are also utilized to delete replicated messages that have already been delivered. This prioritized delivery scheme has been shown to timely deliver packets at vehicular speeds and with tight constraints on buffer spaces. In this case another version of a "contact graph" is considered with edge weights given by the probability that two nodes meet.

The RAPID protocol also takes a micro time scale approach and defines utility, calculated at the packet level, as a function of the inter-meeting time between nodes. Replication decisions are based on optimizing the measured utility under finite buffer constraints. The proposed approach directly considers the impact of replication on network performance. Using testbed traces from a vehicular network, RAPID exhibits performance improvements in terms of average delay and delivery rate over Spray and Wait, PRoPHET, and MaxProp. Here the weights of a contact graph representation would be set by the RAPID utility.

## THE SOCIAL GRAPH FOR OPPORTUNISTIC COMMUNICATION

Many papers in the literature have shown that the random mobility model is not a realistic assumption, and that users tend to have mobility patterns influenced by their social relationships and/or by their attraction to physical places that have special meaning with respect to their social behavior. Routing approaches with the addition of a social graph provide performance improvements over state-of-the-art DTN routing protocols that are not explicitly social.

The links in the social graph we consider may be known a priori or inferred from the frequency of observed contacts. Conti and Kumar identify two social levels in the opportunistic environment: the virtual social network and the electronic social network [13]. Links in the electronic social network depend on the physical properties of the network. Within our graph definitions, the electronic social network could be defined based on analysis of the contact graph. The virtual social network, however, is seen as an overlay network; information about the interpersonal relationships of users can be gained from this level.

We include in our social graph category any protocol that uses information extracted from *a* social layer. A social layer could be inferred from shared context, identified by the application of social network analysis on the contact graph, or constructed from interpersonal relationships available to the network designer. In the remainder of this section, we will present our social graph definition with respect to the virtual social network and the electronic social network [13]. Discussed below are HiBop [4], SimBetTS [6], and BUBBLE [5], which are among the most widely referenced social-based routing protocols.

*Virtual Social Network*

Social-based routing solutions make decisions based on information from a social graph. While social relationships may form due to repeated contact, an interpersonal relationship may exist that is not evident from the contact graph. A physical link may not exist at each instance of time, however, future contact is expected based on the interpersonal relationships of users. This social component may develop through repeated contact, shared interests, geographic preferences, and/or external influences like hierarchal structures. Social-based protocols leverage the relationships identified through these commonalities at the virtual social layer.

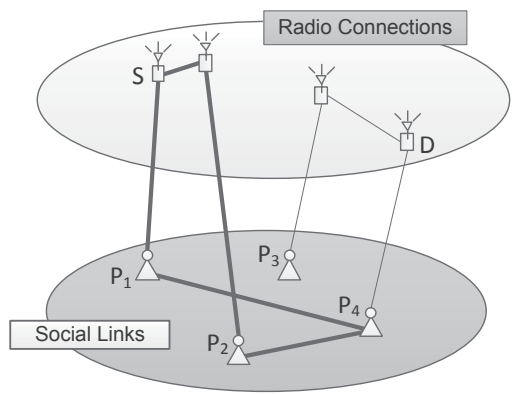

Fig. 3. Underlying social component of wireless links

The integration of two *distinct* communication and social layers in [14] is shown to increase routing robustness at the expense of added delay due to the higher cost assigned to social links. The communication layer is composed of links between devices, and the social layer is formed by equipment owners. Fig. 3 illustrates the interaction of DTN links and the underlying social connections of the users carrying devices. If links at the device level do not exist, an alternate route can be traced through the social layer: $S$ can transmit the message to the device of user $P_2$ and then $P_2$ can carry the message to $P_4$; or $P_1$ can deliver the message directly to $P_4$ through social interaction. The investigated QoS routing approach leverages these social interactions by including the social links as feasible paths and assigning them heavier weights for the routing decision. When using only the contact graph, routes exploiting virtual social links may have never been considered as valid despite the improvement to overall robustness, at the expense of added delay.

Often, especially in opportunistic networks, the communication and social layers are not disjoint. Social relationships impact human mobility and as a result the available connections. HiBop considers mobility with context information from the virtual social layer to construct a type of social graph to predict future connections.

*HiBop:* HiBOp (History Based Opportunistic Routing) [4] uses past and current context information like shared attributes and history of encounters to calculate delivery probabilities. The context information may describe the user's environment and capture social relationships among nodes. The message is transferred if the encountered node's delivery probability for the destination is greater than the current node. The source nodes may replicate messages and inject several copies into the network. When compared to Epidemic and PRoPHET in community-based mobility simulations, HiBOp reduces the consumption of resources and message loss rate for limited buffer scenarios. Delay, however, is shown to increase with HiBOp.

*Electronic Social Network*

The social graph can be viewed as an extension of the contact graph. Knowing $G(contacts)$ (see Fig. 2) different rules may be considered to extract the social graph connections of $SG$. In our example, we assign a social link if nodes meet more than $\frac{1}{3}$ of the time. In the social graph $SG$, $A$ and $C$ belong to two separate clusters and can never communicate. As a consequence, a challenging part of social-based routing design is concerned with the issue of learning/inferring the underlying social interactions from contact history. The identification of central nodes, which connect communities, is also fundamental to this approach.

By representing links from a graph-theoretic perspective, a node's role in an hoc network can be identified through social network analysis. Social network analysis (SNA) examines the relationships between users to identify patterns and quantify network structure. In SNA, the user is not considered as an individual. Instead, the users and their ties (represented by edges) are viewed together as an entity [15]. A goal of SNA is to model connections and to create a structural picture of the network.

Metrics to characterize the social graph (or contact graph) include *degree centrality*, *betweenness centrality*, and *similarity*. The *degree* of a node is the number of adjacent connected nodes [15]. *Betweenness centrality* can be easily described as the number of times a node lies on the shortest path between a source and destination in the network [5]. The *similarity* of two nodes can be measured by the number of shared neighbors [6]. A more recent measure to quantify centrality in a sensor network is defined in [16] as the $\mu$-*power community index*. This metric considers the degree of a node as well as the degree of the node's $\mu$-hop neighbors. As the research on applying social network theory to DTNs is still in its infancy, a question to be answered is: *are there more suitable SNA metrics to characterize the structure of DTNs?*

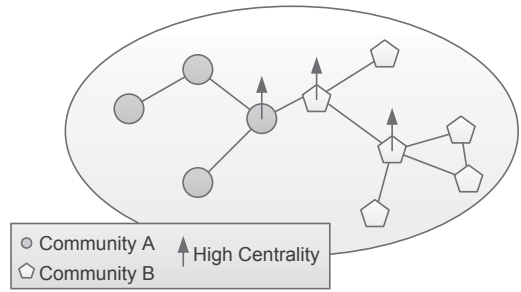

Fig. 4. SNA Metrics identifying popular nodes

Fig. 4 illustrates the use of SNA to characterize the sample network formed by nodes in two communities. The nodes presented with an up arrow have high betweenness centrality and bridge clustered nodes. Nodes with relatively high betweenness and degree centrality measures are seen as nodes with high popularity. The social structure assessed through SNA and subsequent identification of popular nodes will differ depending on the construction of the social graph.

The relative popularity of a node is based on the number of connections and its ability to bridge the partitioned network. Node $E$ in Fig. 2 is a well-connected node with high popularity at $TE3$. We define here the concepts of *static popularity* and *dynamic popularity*. *Static popularity* describes the connectivity of nodes in a predefined social network at the virtual level. *Dynamic popularity* refers to the social structure inferred from the observation of physical links over time. Differences may exist between static and dynamic popularity; thus impacting the identification of highly connected (or isolated) nodes. The characterization of the social network influences routing protocol performance. As social-based routing protocols develop, continued examination of static vs. dynamic popularity is crucial to accurately predicting performance.

The use of SNA metrics to model the wireless network can be further extended to define routing protocols. Efficient routing schemes have transitioned from capturing the frequency of pairwise meetings on the contact graph to utilizing a global view of the electronic social graph as with SimBetTS and BUBBLE.

*SimBetTS:* SimBetTS is the next iteration of Daly and Haahr's SimBet algorithm [6]. The calculations of similarity and betweenness centrality using ego networks allow for a distributed implementation. While the sociocentric network is defined based on global information, egocentric calculations can be performed locally at the ego node. The betweenness aspect of the SimBetTS utility measures the bridging capability of nodes, and similarity identifies nodes socially similar to the destination. SimBetTS utilizes the bridging capability of weak ties and the strong relationships that bind clusters.

SimBetTS also includes tie strength in the utility calculation. Tie strength is seen as an indicator of link availability and is measured by the frequency of encounters, the duration of encounters, and how recently the contact occurred. A replication component is also included in SimBetTS to increase the likelihood of message delivery. While the betweenness measure alone yields the best delivery results, the combined utility, SimBetTS, prevents the overloading of highly central nodes. Balancing the use of popular nodes is ideal from a multi-objective perspective. Message delivery for SimBetTS outperforms PRoPHET and is close to Epidemic with less overhead.

*Bubble Rap:* Following on the LABEL approach, which was the first protocol to demonstrate that incorporating a community affiliation label will improve forwarding performance, BUBBLE expands on this idea by using community affiliation labels with betweenness centrality measures to forward messages [5]. A minimum of two centrality measures are calculated per node based on the node's global popularity in the whole network and local popularity within its community or communities. The algorithm calls for a message to be transferred to nodes with higher global rankings (centrality) until the carrier encounters a node with the same community label as the destination node. The message is then forwarded to nodes with higher local rankings until successful delivery. This approach prevents messages from getting stuck at a node with a high global rank, but with little or no affiliation with the destination community.

Community detection and centrality estimation influence the design of BUBBLE. Centralized and distributed degree and betweenness measures impact the protocol performance. Through simulations, the centralized BUBBLE approach is shown to provide performance improvements in terms of resource utilization compared to flooding and PRoPHET. A modified version of BUBBLE deletes the message from the buffer of the original carrier once the message is transferred to the destination community. Results show that decreasing the number of copies (further reducing the cost) does not negatively impact the delivery ratio for the cases studied.

## Conclusions and Open Issues

Delay-tolerant networks are formed due to partitions in the wireless network. Connectivity exists within clusters, but protocols rely on mobile nodes to route messages between communities. Protocols evolved from flooding all nodes in the network to carefully identifying bridge nodes to carry and forward data. Improvements to benchmark schemes like Epidemic, PRoPHET, and Spray and Wait have produced performance enhancements in the form of reduced communication costs and comparable delivery ratios. Predicting and exploiting pairwise contacts has led to the extension of social network theory to wireless networks.

In this work we present DTN routing protocols based on the network graph considered; classifications are based on the wireless graph, contact graph, and social graph. Table I summarizes the described protocols and identifies the corresponding network graph. Solutions using the social graph form a new class of routing protocols well suited for opportunistic networks. Despite recent advances, there are still opportunities for development. We will now discuss the overarching open issues for intermittently connected networks and transition to challenges specific to social network-based solutions.

DTNs may form, in some cases, between different types of nodes operating with incompatible hardware and software. Interoperability is an ongoing issue for these heterogeneous networks of dissimilar nodes. The Delay Tolerant Networking Research Group[1] is tasked with addressing interconnection in such networks.

Included in the list of open research topics for all intermittently connected networks are the issues of security and the nodes' possible selfish behavior. How can trust be measured and propagated through the network? Are all nodes willing to act as relays? Can privacy be maintained and to what extent? What incentives can entice selfish nodes to participate in forwarding? The integration of a social component may be central to overcoming these challenges.

In terms of DTN routing performance, representative mobility models are needed for accurate protocol evaluation. Also, current approaches typically assume perfect transmission during pairwise contact. The incorporation of interference and bandwidth limitations will provide tighter bounds on expected performance.

Likewise, a multi-objective approach which aims to concurrently optimize criteria may provide significant insight into performance tradeoffs. As DTN protocols continue to evolve, a balance should be reached between robust delivery, expected delay, total energy consumption, and buffer utilization.

The identification of popular, well-connected nodes is fundamental to the social-based approach. However, protocols which overuse these nodes may experience a degradation in performance. Intuitively, message delivery should increase, but the overall delivery ratios will likely decrease in practice due to the limited capacity of finite buffers. The expected delay may increase as well due to contention at highly central nodes. Studies suggest that the integration of some level of randomness into protocol design may benefit performance.

The underlying traffic patterns and sociability of nodes also relate to protocol performance. The aggregation windows used to define the contact or social graphs must be finely tuned. Concepts from machine learning or signal processing may aid in this effort.

While the literature contains a wealth of information regarding an inferred social structure, there is more work to be done to incorporate a predefined hierarchy. Inconsistencies exist between static and dynamic popularity. Understanding their performance differences should be further explored.

Social-based routing approaches may bring DTN performance closer to optimal bounds, but distributed implementations need to be further developed before such performance can be realized in practice. Haggle[2] and SocialNets[3] are two projects which aim to address opportunistic networking among deployed devices. A theme of this type of work is that limited connectivity may not always be a challenge to overcome, but instead, an opportunity to construct a new type of network for pervasive computing [13]. With the advent of these human-centric networks, opportunistic networking research will surely continue to develop.

---

[1]http://www.dtnrg.org/
[2]http://www.haggleproject.org/
[3]http://www.social-nets.eu/

TABLE I
DTN State of the Art Protocol Overview

| Protocol | Year | Network Graph | Description |
|---|---|---|---|
| Epidemic | 2000 | Wireless | Assumes disconnection and relies on mobility to forward. Random pair-wise exchange of messages (anti-entropy sessions). Aims to minimize number of transmissions by imposing a max hop count and a bound on buffer space. |
| PRoPHET | 2003 | Contact | Probabilistic Routing Protocol using History of Encounters and Transitivity - Based on assumption node mobility is not random. Forwards message if delivery predictability is higher at other node. Based on number of encounters, age of encounters, and transitive property. |
| Spray and Wait | 2005 | Wireless | Distributes L copies of a message into the network. Once copies are forwarded, carriers hold until they reach the destination. |
| MaxProp | 2006 | Contact | Prioritizes packets based on delivery likelihood at destination and total hop count. Complementary mechanisms like acknowledgements further increase delivery and decrease latency. |
| RAPID | 2007 | Contact | Replicates a packet based on a routing metric and per-packet utility measure. Control channel allows for the exchange of network state information including acknowledgements. |
| HiBOp | 2007 | Social | History Based Opportunistic Routing - Identifies appropriate carriers based on shared context with destination. Eliminates unnecessary replication to disjoint clusters. |
| SimBet SimBetTS | 2007 2009 | Social | Utility based on similarity and betweenness measures. SimBetTS described in 2009 extended utility to include tie strength. At encounter if node has higher utility for a given destination, messages are exchanged and removed from queue based on replication definition. |
| BUBBLE BUBBLE-B | 2008 2010 | Social | Utilizes community and rank information. Ranks are based on local and global betweenness centrality values. Forward if encountered node has higher global rank then higher local rank once reach community of destination. For BUBBLE-B, described in 2010, deletes from original buffer once it reaches community of destination. |

Social networks and opportunistic networks are intertwined due to the underlying human component. For other DTNs with random mobility or predictable schedules, social-based solutions may not be the best answer. All in all, social networking for opportunistic communication is an interesting research area and worth pursuing for most intermittently connected ad hoc networks.

## Acknowledgment

The authors would like to thank the reviewers for their insightful feedback and valuable suggestions. Additionally, the first author is grateful for her research experiences at BAE Systems and MIT Lincoln Laboratory. Special thanks to Dr. Reza Ghanadan and Dr. Ladan Gharai for introducing the topic of social networking for wireless communications at BAE Systems. Also, special thanks to Dr. Wayne Phoel, Dr. Jeffrey Wysocarski, and Dr. Michele Schuman of MIT Lincoln Laboratory for their guidance and the opportunity to examine the relevance of social-based routing to the first author's Ph.D. research.

## References

[1] K. Fall, "A Delay-Tolerant Network Architecture for Challenged Internets," in *Proceedings of the 2003 conference on Applications, technologies, architectures, and protocols for computer communications*, ser. SIGCOMM '03. New York, NY, USA: ACM, 2003, pp. 27–34.
[2] S. Jain, K. Fall, and R. Patra, "Routing in a Delay Tolerant Network," *SIGCOMM Comput. Commun. Rev.*, vol. 34, pp. 145–158, Aug. 2004.
[3] M. Wang and K. Nahrstedt, "Social Structure Based Routing of Intermittently Connected Network Using Contact Information," in *Proc. IEEE MILCOM*, San Diego, California, USA, Nov. 2008, pp. 1–7.
[4] C. Boldrini, M. Conti, and A. Passarella, "Exploiting users' social relations to forward data in opportunistic networks: The HiBOp solution," *Pervasive and Mobile Computing*, vol. 4, no. 5, pp. 633–657, 2008.
[5] P. Hui, J. Crowcroft, and E. Yoneki, "BUBBLE Rap: Social-Based Forwarding in Delay Tolerant Networks," *IEEE Transactions on Mobile Computing*, vol. 10, no. 11, pp. 1576–1589, Nov. 2011.
[6] E. Daly and M. Haahr, "Social Network Analysis for Information Flow in Disconnected Delay-Tolerant MANETs," *IEEE Transactions on Mobile Computing*, vol. 8, no. 5, pp. 606–621, May 2009.
[7] G. Ding and B. Bhargava, "Peer-to-peer File-sharing over Mobile Ad Hoc Networks," in *Proc. IEEE PerComW'04*, Orlando, Florida, USA, Mar. 2004, pp. 104–108.
[8] A. Vahdat and D. Becker, "Epidemic Routing for Partially-Connected Ad Hoc Networks," Duke University, Tech. Rep. CS-2000-06, 2000.

[9] T. Spyropoulos, K. Psounis, and C. S. Raghavendra, "Efficient Routing in Intermittently Connected Mobile Networks: The Multiple-Copy Case," *IEEE/ACM Trans. Netw.*, vol. 16, pp. 77–90, Feb. 2008.

[10] A. Lindgren, A. Doria, and O. Scheln, "Probabilistic Routing in Intermittently Connected Networks," in *Service Assurance with Partial and Intermittent Resources*, ser. Lecture Notes in Computer Science, P. Dini, P. Lorenz, and J. N. d. Souza, Eds.  Springer Berlin / Heidelberg, 2004, vol. 3126, pp. 239–254.

[11] J. Burgess, B. Gallagher, D. Jensen, and B. N. Levine, "MaxProp: Routing for Vehicle-Based Disruption-Tolerant Networks," in *Proc. IEEE INFOCOM*, Barcelona, Spain, Apr. 2006.

[12] A. Balasubramanian, B. Levine, and A. Venkataramani, "DTN Routing as a Resource Allocation Problem," *ACM SIGCOMM Comput. Commun. Rev.*, vol. 37, pp. 373–384, Aug. 2007.

[13] M. Conti and M. Kumar, "Opportunities in Opportunistic Computing," *Computer*, vol. 43, no. 1, pp. 42–50, Jan. 2010.

[14] M. Bigrigg, K. Carley, K. Manousakis, and A. McAuley, "Routing Through an Integrated Communication and Social Network," in *Proc. IEEE MILCOM*, Boston, Massachusetts, USA, Oct. 2009, pp. 1–7.

[15] S. Wasswerman and K. Faust, *Social Network Analysis: Methods and Applications*.  Cambridge, United Kingdom: Cambridge University Press, 1994.

[16] A. Papadimitriou, D. Katsaros, and Y. Manolopoulos, "Social Network Analysis and Its Applications in Wireless Sensor and Vehicular Networks," in *Next Generation Society. Technological and Legal Issues*, ser. Lecture Notes of the Institute for Computer Sciences, Social Informatics and Telecommunications Engineering, O. Akan, P. Bellavista, J. Cao, F. Dressler, D. Ferrari, M. Gerla, H. Kobayashi, S. Palazzo, S. Sahni, X. S. Shen, M. Stan, J. Xiaohua, A. Zomaya, G. Coulson, A. B. Sideridis, and C. Z. Patrikakis, Eds.  Springer Berlin Heidelberg, 2010, vol. 26, pp. 411–420.

**Mary R. Schurgot** is a Ph.D. candidate in electrical engineering at Stevens Institute of Technology. In 2006, she completed her B.E. degree and later in 2008 her M.E.E., both from Stevens. She currently works as a Cybersecurity Research Engineer at LGS Innovations. She has previously contributed to research projects while with BAE Systems, Inria, and MIT Lincoln Laboratory. Her current research interests are focused in the areas of wireless network optimization and military communications.

**Cristina Comaniciu** received the M.S. degree in electronics from the Polytechnic University of Bucharest in 1993, and the Ph.D. degree in electrical and computer engineering from Rutgers University in 2002. From 2002 to 2003 she was a postdoctoral fellow with the Department of Electrical Engineering, Princeton University. Since August 2003, she is with Stevens Institute of Technology, Department of Electrical and Computer Engineering, where she is now an Associate Professor. In Fall 2011 she was a visiting faculty fellow with the Department of Electrical Engineering, Princeton University. She served as an associate editor for the IEEE Communication Letters Journal (2007-2011), and she is currently a member of the IEEE Interest Group on Green Multimedia Communications. Professor Comaniciu is a recipient of the 2007 IEEE Marconi Best Paper Prize Award in Wireless Communications and of the 2012 Rutgers School of Engineering Distinguished Young Alumnus Medal of Excellence. She is a coauthor of the book "Wireless Networks: Multiuser Detection in Cross-Layer Design," Springer, NY.

**Katia Jaffrès-Runser** received both a Dipl. Ing. (M.Sc.) in Telecommunications and a DEA (M.Sc) in Medical Imaging in 2002 and a Ph.D. in Computer Science in 2005 from the National Institute of Applied Sciences (INSA), Lyon, France. From 2002 to 2005 she was with Inria, participating in the ARES project while working towards her Ph.D. thesis. In 2006, she joined the Stevens Institute of Technology, Hoboken, NJ, USA, as a post-doctoral researcher. She is the recipient of a three-year Marie-Curie OIF fellowship from the European Union to pursue her work from 2007 to 2010 in both Stevens Institute of Technology and INSA Lyon on wireless networks modeling and multiobjective optimization. In 2011, she participated in the GreenTouch consortium as a delegate from Inria. Since September 2011, she joined the University of Toulouse - ENSEEIHT as a Maître de Conférences (Associate Professor), working at IRIT laboratory on hybrid embedded networks optimization.